# Evidence of Inhomogeneous Superconductivity in FeTe$_{1-x}$Se$_x$ Thin Film Using Scotch-Tape Method

H. Okazaki,[1] T. Watanabe,[1,2] T. Yamaguchi,[1] Y. Kawasaki,[1,2] K. Deguchi,[1,2] S. Demura,[1,2] T. Ozaki,[1] S. J. Denholme, Y. Mizuguchi,[3] H. Takeya,[1] Y. Takano[1]

[1]National Institute for Materials Science, 1-2-1 Sengen, Tsukuba, Ibaraki 305-0047, Japan

[2]Graduate School of Pure and Applied Sciences, University of Tsukuba, 1-1-1 Tennodai, Tsukuba, Ibaraki 305-8577 Japan

[3]Department of Electrical and Electronic Engineering, Tokyo Metropolitan University, 1-1, Minami-osawa, Hachioji, 192-0397, Japan

**Abstract**

We have fabricated thin films of FeTe$_{1-x}$Se$_x$ using a scotch-tape method. The superconductivities of the thin films are different from each other although these films were fabricated from the same bulk sample. The result clearly presents the inhomogeneous superconductivity in FeTe$_{1-x}$Se$_x$. The difference comes from inhomogeneity due to the excess Fe concentration. The resistivity of a thin film with low excess Fe shows good superconductivity with the sharp superconducting-transition width and more isotropic superconductivity.

## 1. Introduction

Fe-based superconductors discovered in 2008 [1] have stimulated fundamental discussion on the mechanism of superconductivity. Among the Fe-based superconductors, 11 type compounds have the simplest structure consisting only of superconducting layers. On the other hand, the other Fe-based superconductors (111 type, 122 type, 1111 type, and so on) have a stacking structure with both superconducting layers and block layers. Thus, 11 type superconductors have been extensively studied in order to understand the mechanism of Fe-based superconductivity. The 11 type superconductor FeSe exhibits superconductivity with $T_c$ = 13K [2] and its $T_c$ reaches 37K under high pressure (4-6 GPa) [3,4]. The pressure dependence of $T_c$ relates to the anion height from the Fe layers [2], and $T_c$ of FeSe increases with decreasing anion (Se) height by high pressure. Thus, 11 type superconductors have the potential of achieving a high $T_c$. Additionally, $T_c$ of $K_{0.8}Fe_2Se_2$, composed of K intercalation between FeSe layers, also shows an abrupt increase above 30 K [5]. For $(LiNH_2)_xFe_2Se_2$, the superconductivity exhibits a higher $T_c$ of 43 K [6]. With an increase in the $c$ lattice parameter and the separation between Fe-chalcogen layers, $T_c$ increases. Therefore, a change of Fe-chalcogen layer has a great impact on superconductivity. Recently, it was reported that a FeSe monolayer synthesized by MBE method shows superconductivity at 53 K [7]. σ-GGA (spin-dependent generalized gradient approximation) calculations have predicted that a Dirac-cone state appears in a FeSe monolayer [8]. The Dirac-cone states behind the superconducting state may substantially raise its $T_c$ [9]. Furthermore, there are several reports where the $T_c$ of superconducting ultrathin films become higher than that of the bulk [10,11]. Thus, an ultrathin film of 11 type compounds may show an abrupt increase in $T_c$.

Recently, it was reported that thin films of the 11 type superconductor $FeTe_{1-x}S_x$ can be fabricated by the scotch-tape method [12], which is a powerful technique to easily prepare an ultrathin film such as graphene [13]. The structure of 11 type compounds is similar to that of graphite since each layer is connected by van-der-Waals forces. In addition, 11 type superconductors have no insulating block layer. Thus, the 11 type superconductors are ideal candidates for the fabrication of a superconducting monolayer film by the scotch-tape method. If we could obtain a monolayer or ultrathin film, it will give us a greater understanding of the superconducting mechanism. Additionally, a property which has not been observed for the bulk yet may appear by the thinning. Thus, we fabricated the thin films and measured the resistivities.

## 2. Experimental

Single crystals of $FeTe_{0.65}Se_{0.35}$ were synthesized by a self-flux method. Powders of Fe (99.9%), Te (99.999%), and Se (99.9999%) were sealed into evacuated quartz tube with the desired molar ratio. Since the tube often cracks during the cooling, it was sealed into a larger evacuated tube. The doubly-sealed materials were heated at 1100 °C for 20 hours and then cooled about -2 °C/h down to 650 °C followed by furnace cooling. The obtained single crystal was placed on scotch tape, and cleaved several times. The scotch tape with the $FeTe_{0.65}Se_{0.35}$ flakes were attached to a silicon substrate with a silicon dioxide layer on the top surface, and pressed to enhance the bonding between the crystals and the substrate by van-der-Waals forces. By this process [12], thin single crystals were left on the substrate.

An optical microscope and an atomic force microscope (AFM) were used to investigate the thin films fabricated by the scotch-tape method. We measured the resistivity of the thin films by four-terminal circuit fabricated using electron-beam lithography. In the process of electron-beam lithography, a resist was first spin-coated onto the substrates. The resist underwent a specially designed electron-beam scan routine in order to write electrode patterns. The irradiated area was removed by a developer, and then the electrode was deposited. The residual resist was finally taken off. By this technique, we prepared the four-terminal electrodes. The width and separation of the terminals are 2 and 3 μm, respectively.

## 3. Results and discussion

Figure 1(a)-(c) and (d)-(f) show optical microscopic and AFM images of $FeTe_{0.65}Se_{0.35}$ thin films fabricated by the scotch-tape method. From the height profile at the line indicated in Fig. 1(d)-(f), we observed that the surfaces of these films are flat, as shown in Figs. 1(g)-(i). These results indicate that a thin film with a flat surface can easily be obtained using the scotch-tape method. The thicknesses of these films were estimated to be 15, 65, and 100 nm, respectively. From the comparison between the optical and AFM images, we found that the colors of the thin films are gradually paled out with decreasing thicknesses.

The temperature dependence of resistivities of these films with thicknesses of 15, 65, 100 nm, and the bulk are different from each other, as shown in Fig. 2. The same onset superconducting transition temperature $T_c^{on}$ of 15.0 K is observed for 65, 100 nm, and the bulk. However, the thin film with 15nm thickness does not show a superconducting transition. The transition widths ($\Delta T_c$) are $\Delta T_c$ = 4.6 K for 65 nm, $\Delta T_c$ = 1.0 K for 100 nm, and $\Delta T_c$ = 6.5 K for the bulk, respectively. The thin film with 100 nm thickness clearly shows a sharper superconducting transition than other films and the bulk. These results clearly show the inhomogeneous superconductivity in $FeTe_{1-x}Se_x$ although these films were extracted from the same bulk sample. Furthermore, the normal-state resistivities above $T_c$ show not only metallic properties but also semiconductivity. The normal-state resistivities for 65 and 100 nm slowly decreases with decreasing temperature while that for 15 nm and the bulk increases, as shown in Fig. 2(b). The broadening superconducting transition and the change from metallic to semiconducting behavior of each film are similar to the change of $Fe_{1+d}Te_{1-x}Se_x$ with increasing excess Fe composition [14,15]. Therefore, the increase of the resistivity corresponds to weak charge carrier localization due to the excess Fe, which results in the broadening of superconducting transition for 65 nm and the bulk. The existence of the excess Fe makes the crystal inhomogeneous. Thus, the result for 100 nm, which shows metallic resistivity with sharpest $\Delta T_c$, suggests that a homogeneous crystal with low excess Fe, originally part of the bulk, is cleaved and remains on the surface of the substrate incidentally. This indicates that a homogeneous crystal with low excess Fe can be obtained using the scotch-tape method and shows ideal superconductivity. On the other hand, the result for 65nm implies that a part of the crystal with some excess Fe remains on the substrate and its superconductivity is inhibited by the excess Fe. For 15nm, a part of the crystal with a large excess of Fe remains and the superconductivity is completely suppressed and changes from metallic character to semiconductivity.

Additionally, we determined the upper critical field $H_{c2}$ ($T$) curves of the thin film with 100 nm thickness for $H//c$ axis and the $H//ab$ plane from the temperature dependence of the resistivity under a magnetic field. The temperature dependence of the resistivity under a magnetic field is summarized in Fig. 3 for (a) the $H//c$ axis and (b) $H//ab$ plane. The temperature at which the resistivity reached zero was systematically suppressed with increasing magnetic field. We estimated $H_{c2}$ ($T$) from the criterion of 90% of normal-state resistivity. From the comparison of $H_{c2}$ ($T$) for the thin film with bulk, $H_{c2}$ ($T$) curves of the thin film shift toward higher field than those of bulk. Moreover, we calculated the $H_{c2}^{ab}$ (0) and $H_{c2}^{c}$ (0) using the data from 0 to 7T and Werthamer-Helfand-Hohenberg (WHH) theory [16]: $H_{c2}^{ab}$ (0) = 134.3 T and $H_{c2}^{c}$ (0) = 90.6 T for the thin film, and $H_{c2}^{ab}$ (0) = 87.9 T and $H_{c2}^{c}$ (0) = 39.2 T for the bulk. We also estimated the irreversibility field from the criterion of 10% of normal-state resistivity, as shown in Fig 3(d): $H_{irr}^{ab}$ (0) = 102.6 T and $H_{irr}^{c}$ (0) = 45.6 T for the film, and $H_{irr}^{ab}$ (0) = 53.9 T and $H_{irr}^{c}$ (0) = 27.7 T for the bulk. The $H_{irr}$ for the film are much larger than that of the bulk and the previous reports [17,18]. The anisotropy coefficients $\gamma$ determined from $\gamma = H_{c2}^{ab}$ (0)/ $H_{c2}^{c}$ (0) were estimated to be 1.5 for the thin film and 2.2 for the bulk, respectively. The anisotropy coefficient $\gamma$ of the thin film shows more isotropic behavior than that of the bulk. We found that the film with the lowest $\Delta T_c$ exhibits more isotropic $\gamma$, high $H_{c2}$ and $H_{irr}$. In previous reports, the bulk sample with low excess Fe shows low $\Delta T_c$ than high excess Fe [14]. Excess Fe induces the broadening $\Delta T_c$, worse superconducting properties, and more anisotropic behavior than ideal $FeTe_{1-x}Se_x$. Therefore, our results indicate that ideal $FeTe_{1-x}Se_x$ bulk should exhibit high superconducting performance with isotropic superconductivity and has an advantage to application.

## 4. Conclusion

We have fabricated thin films of $FeTe_{0.65}Se_{0.35}$ using the scotch-tape method, and measured the resistivities of the thin films and the bulk sample. The different superconductivities in the thin films and the bulk were observed in spite of the films extracted from same bulk. The result clearly shows $FeTe_{1-x}Se_x$ crystal is inhomogeneous by location. The inhomogeneity comes from the excess Fe concentration by location. The resistivity of a thin film with low excess Fe shows that the excess Fe induces the lowest $\Delta T_c$, high $H_{c2}$ and $H_{irr}$.

**Acknowledgements**

We thank A. Kanda, and H. Tomori for valuable discussion about scotch-tape method. This work was supported partly by JST-TRiP and JST-EU-Japan SC.

**Figure captions**

Fig. 1. (a)-(c) Optical images of FeTe$_{0.65}$Se$_{0.35}$ thin film fabricated by the scotch-tape method corresponding to the thickness of (a) 15 nm, (b) 65nm, and (c) 100 nm, respectively. (d)-(f) AFM images scanning the thin films in (a)-(c). (g)-(i) The height profile of these films corresponding to the lines in (d)-(f).

Fig. 2. Temperature dependence of resistivity of the thin films with thickness of 15, 65, 100 nm, and the bulk for (a) 5-20 K and (b) 5-300 K. Filled circles, triangles, squares, and open circles

correspond to the resistivity of the thin film with 15, 65, 100 nm, and the bulk, respectively.

Fig. 3 The temperature dependence of the resistivity of 100nm thickness from 0 to 7 T along (a) the $H//c$ axis and (b) $H//ab$ plane. (c) The $H_{c2}$ ($T$) curves of the thin film and the bulk sample for the $H//c$ axis and the $H//ab$ plane. The values of $H_{c2}$ ($T$) are determined by criterion of 90% of normal-state resistivity. Circles and triangles correspond to the $H_{c2}$ ($T$) of the thin film and the bulk, respectively.

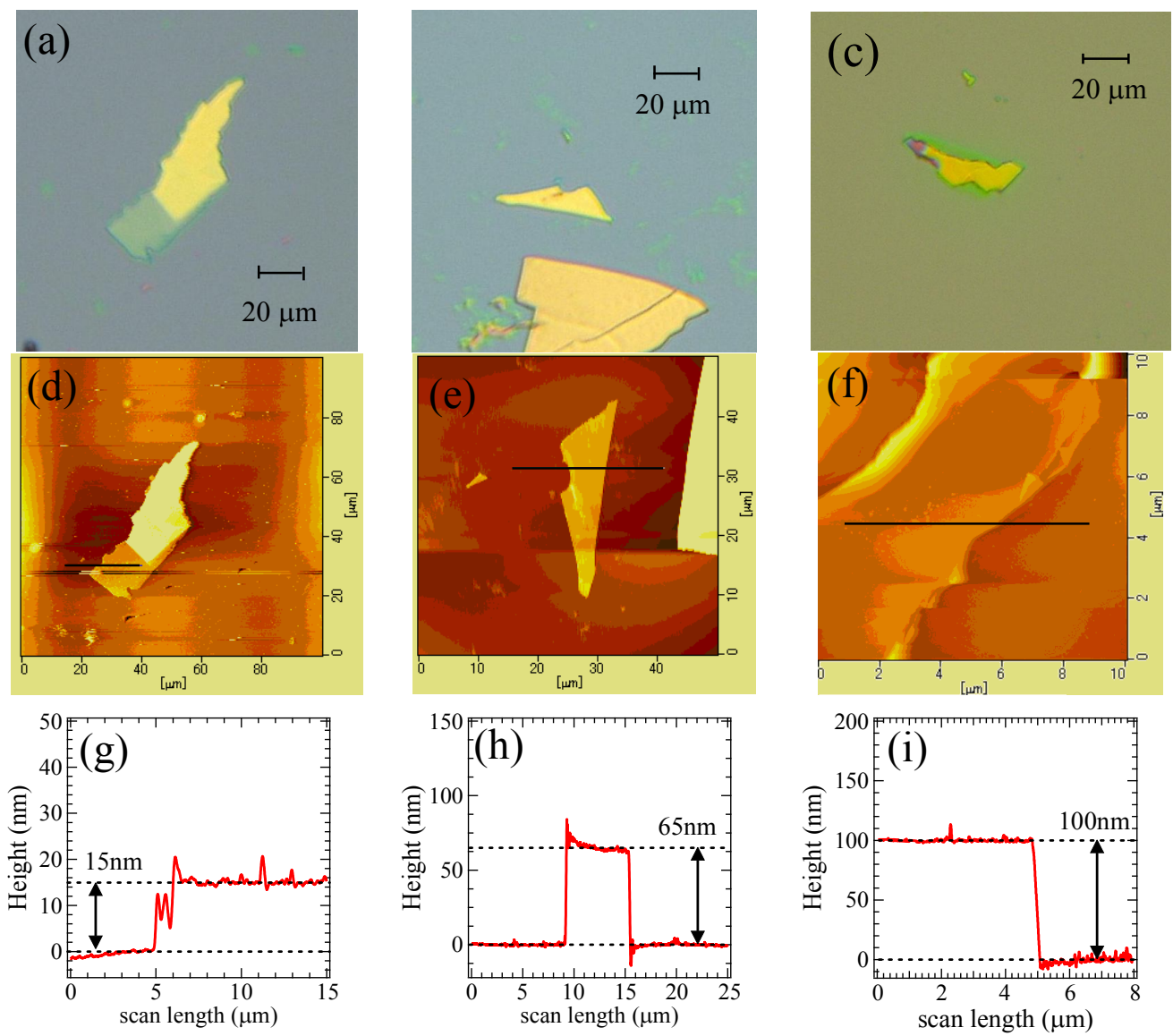

Fig. 1. H. Okazaki

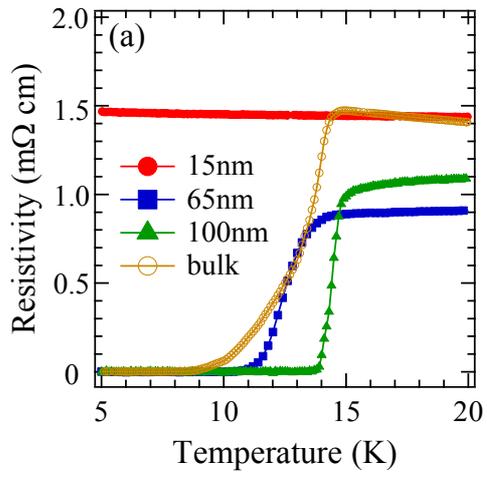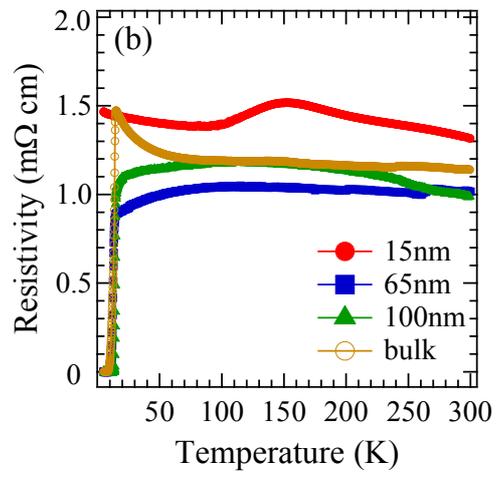

Fig.2. H. Okazaki

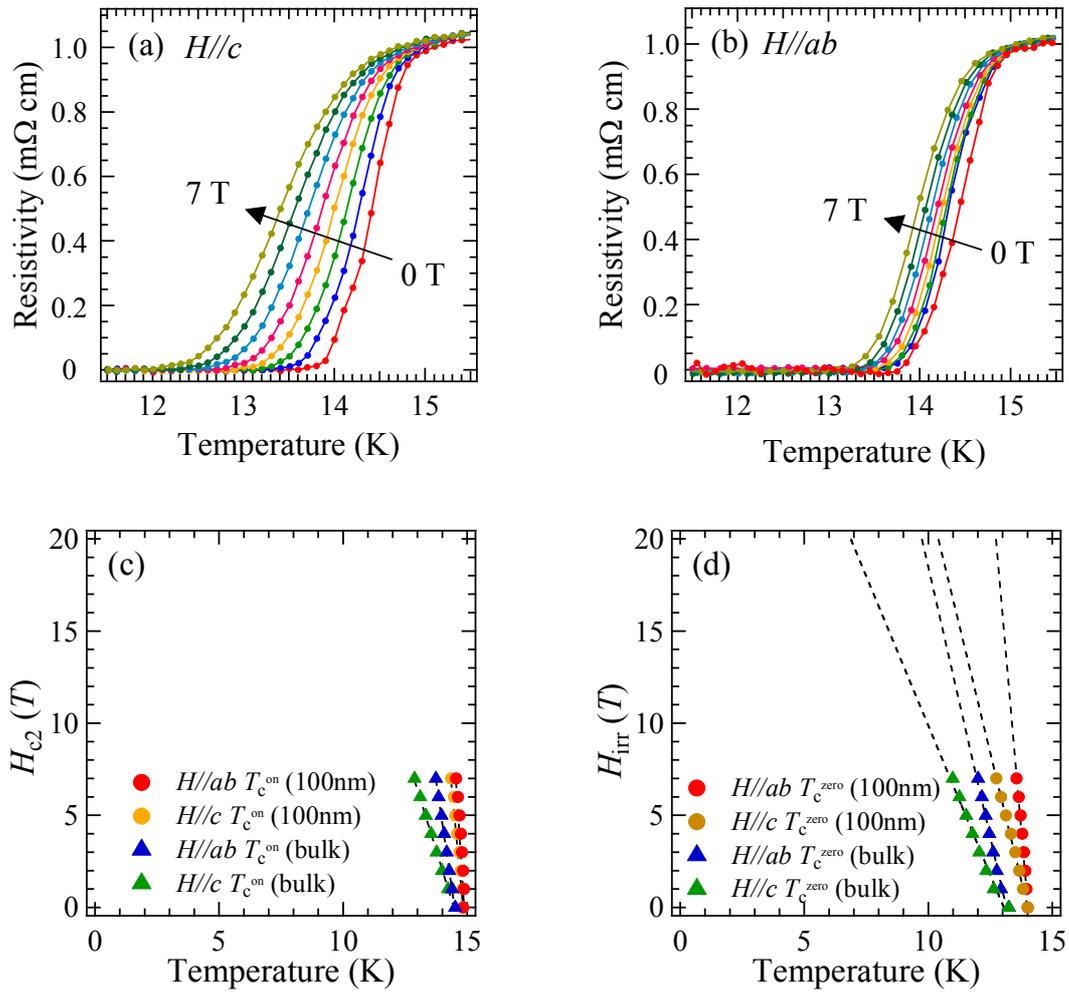

Fig. 3. H. Okazaki